\documentclass[aps,prl,reprint,superscriptaddress,nofootinbib,floatfix]{revtex4-2}
\usepackage{amsmath,amssymb,bm,graphicx,booktabs,array,multirow}
\usepackage[hidelinks]{hyperref}

\newcommand{\dmchi}{\Delta m_{21}\equiv m_{\widetilde\chi_2^0}-m_{\widetilde\chi_1^0}}
\newcommand{\oh}{\Omega_{\widetilde\chi}h^2}
\begin{document}

\title{Thermal Higgsino Dark Matter in GmSUGRA for the LZ Event and Solar-Neutrino Constraint}

\author{Imtiaz Khan}
\email{ikhanphys1993@gmail.com}
\affiliation{Department of Physics, Zhejiang Normal University, Jinhua, Zhejiang 321004, China}
\affiliation{Research Center of Astrophysics and Cosmology, Khazar University, Baku, AZ1096, Azerbaijan}
\author{Waqas Ahmed}
\email{waqasmit@hbpu.edu.cn}
\affiliation{Center for Fundamental Physics and School of Artificial Intelligence, Hubei Polytechnic University,  Huangshi 435003, China}
\author{Tianjun Li}
\email{tli@mail.itp.ac.cn}
\affiliation{School of Physics, Henan Normal University, Xinxiang 453007, China}
\author{Shabbar Raza}
\email{shabbar.raza@fuuast.edu.pk}
\affiliation{Department of Physics, Federal Urdu University of Arts, Science and Technology, Karachi 75300, Pakistan}
\author{Ali Muhammad}
\email{alimuhammad@phys.qau.edu.pk}
\affiliation{CAS Key Laboratory of Theoretical Physics, Institute of Theoretical Physics, Chinese Academy of Sciences, Beijing 100190, China}
\affiliation{School of Physical Sciences, University of Chinese Academy of Sciences, No. 19A Yuquan Road, Beijing 100049, China}

\begin{abstract}
We study the thermal Higgsino dark matter in the Minimal Supersymmetric Standard Model (MSSM) with Generalized Minimal Supergravity (GmSUGRA) boundary conditions after the high-recoil event reported by the LUX--ZEPLIN experiment and the subsequent solar-neutrino limits inferred from IceCube data.  The Higgs sector is implemented in a NUHM2-type parametrization with $\mu$ and $m_A$ as inputs, while the supersymmetry breaking soft terms obey the non-universal GmSUGRA relations.  The numerical scan contains a Planck-compatible Higgsino branch near the thermal mass, with the neutral-state splitting extending continuously from the few-hundred-keV region to the MeV scale.  The part of this branch with $\dmchi\simeq0.3$--$0.4$ MeV overlaps the kinematic interval relevant to the LZ event, but is disfavored by the current solar-capture reinterpretations for a full-density thermal Higgsino.  The same high-scale branch persists at larger splittings, including representative spectra with $\dmchi\simeq0.85$, $1.10$, and $1.34$ MeV, beyond the quoted solar-capture limits. Our results demonstrate that the mass and sub-MeV splitting favored by the LZ high-recoil interpretation can be realized dynamically within a GUT-motivated supersymmetric construction, without imposing the neutralino compression as a weak-scale input. The combined data therefore select different parts of a common GmSUGRA Higgsino branch rather than excluding the thermal Higgsino construction itself. 
\end{abstract}

\maketitle

\textit{Introduction.---}
The nature of dark matter (DM) remains one of the open questions in particle physics and cosmology.  Supersymmetric extensions of the Standard Model provide a stable neutral particle when $R$ parity is conserved, and the lightest neutralino can account for the cosmological abundance over several regions of the MSSM parameter space \cite{HaberKane1985,JungmanEtAl1996,MartinPrimer}.  A Higgsino-dominated neutralino is particularly simple: electroweak annihilation and coannihilation select a mass close to the TeV scale when the Higgsino provides essentially all of the thermal relic density.

The extended nuclear-recoil analysis of LUX--ZEPLIN (LZ) reaches recoil energies of about $270$ keV in an exposure of $2.84$ tonne-years \cite{LZ2026HighRecoil}.  One event is consistent with a nuclear recoil at $248\pm23_{\rm stat}\pm23_{\rm sys}$ keV.  The largest local significance among the tested models is $3.4\sigma$, while the global significance is $2.6\sigma$ after the look-elsewhere correction \cite{LZ2026HighRecoil}.  The event therefore provides a useful kinematic target without establishing a DM signal.

Endothermic scattering shifts the recoil spectrum to higher energy.  If the incoming state $\chi_1$ scatters into a heavier state $\chi_2$, the mass difference $\delta=m_{\chi_2}-m_{\chi_1}$ has to be supplied by the kinetic energy of the incident particle.  The two neutral Higgsino states of the MSSM naturally realize this structure because the $Z$ current is dominantly off diagonal in the Majorana mass basis \cite{TuckerSmithWeiner2001,TuckerSmithWeiner2005}.  Recent analyses of the LZ event have therefore considered a thermal Higgsino near $1.1$ TeV with a neutral splitting of a few $10^2$ keV \cite{FreeseTheodosopoulos2026,WuZhangZhu2026,FanReece2026,Yin2026,DuWang2026,RoddEtAl2026,Khan:2026nwp}.  The numerical interval depends on the halo distribution, treatment of the high-energy sideband, and the assumed particle spectrum \cite{RoddEtAl2026,GhoshEtAl2026}.

Beyond the minimal Higgsino interpretation, several recent studies have explored alternative realizations of inelastic dark matter motivated by the LZ high-recoil event. These include extended gauge-sector constructions, electroweak multiplets, ALP- and dark-photon-mediated scenarios, endothermic/exothermic dark matter, extra-dimensional realizations, and isotope-dependent kinematic effects \cite{OkadaSeto2026,SmirnovEtAl2026,YuanEtAl2026,FanHeWangZhao2026,Yamashita2026,AhmedLeontarisDarkDimension2026,AhmedAhmadRehman2026,AhmedLeontarisKK2026}. These studies demonstrate that a few-hundred-keV inelastic splitting can arise from several different microscopic mechanisms, motivating the investigation of whether such a scale can also emerge naturally from a correlated supersymmetric high-scale construction.

The same inelastic interaction is probed differently in the Sun.  Dark matter falling into the solar potential reaches speeds that are not available in terrestrial searches and may upscatter on heavy solar nuclei.  The captured population can annihilate into electroweak final states and produce high-energy neutrinos.  A ten-year IceCube analysis finds no significant excess from the direction of the Sun and places limits on DM annihilation scenarios \cite{IceCubeSolar2025}.  Reinterpretations for an inelastic thermal Higgsino obtain lower limits on the neutral splitting around $0.51$--$0.57$ MeV \cite{PospelovRamani2026,DiMauroShaikh2026,BoseEtAl2026}.  The result is important for the LZ interpretation because the few-hundred-keV region is kinematically accessible in the Sun even when terrestrial scattering is strongly suppressed.  Recent studies also find that variations of the Galactic velocity distribution do not generally remove this tension for the canonical thermal Higgsino \cite{GhoshEtAl2026}.

The question addressed here is different from fitting the LZ event with a freely chosen weak-scale splitting.  We ask whether the LZ-oriented and solar-compatible splittings occur within a common ultraviolet construction once the supersymmetric spectrum is generated from correlated high-scale inputs.  We consider the MSSM with GmSUGRA boundary conditions \cite{LiNanopoulos2010,BalazsEtAl2010,LiRaza2015}.  This framework allows nonuniversal soft terms but relates the electroweak and colored sectors through the underlying GUT-scale structure.  It has been used in a range of neutralino-DM and electroweak-supersymmetry studies \cite{KhanEtAl2024,KhanEtAl2025Light,KhanEtAl2026EWSUSY,KhanEtAlJHEP2026, Khan:2023ryc,AhmedEtAl2018Natural,AhmedEtAl2022MuonG2}.

Our scan gives a TeV Higgsino branch satisfying the relic-density interval while covering both the LZ-oriented splitting and larger MeV-scale splittings.  This separation is the central point of the analysis: the solar-neutrino results restrict the low-splitting part of the branch, whereas the GmSUGRA realization itself remains available at larger $\dmchi$.  We describe below the high-scale construction, numerical procedure, correlations that generate the spectrum, and the relation to the current direct- and indirect-detection constraints.

\textit{GmSUGRA realization.---}
GmSUGRA generalizes universal supergravity boundary conditions while retaining relations among the gauge and gaugino sectors at the unification scale \cite{LiNanopoulos2010,BalazsEtAl2010}.  In the convention used in the scan,
\begin{align}
 \frac{1}{\alpha_2}-\frac{1}{\alpha_3}
 &=k\left(\frac{1}{\alpha_1}-\frac{1}{\alpha_3}\right),\\
 \frac{M_2}{\alpha_2}-\frac{M_3}{\alpha_3}
 &=k\left(\frac{M_1}{\alpha_1}-\frac{M_3}{\alpha_3}\right).
\end{align}
For $k=5/3$ and approximately unified gauge couplings, the gaugino relation becomes
\begin{equation}
 M_3=\frac52 M_1-\frac32 M_2 .
 \label{eq:gaugino}
\end{equation}
Thus the two electroweak gaugino inputs also determine the gluino boundary condition.  This correlation is retained throughout the scan rather than imposing the neutralino splitting as an independent weak-scale parameter.

The scalar sector is also nonuniversal.  The implementation used here satisfies
\begin{align}
 m_{\widetilde Q_i}^2&=\frac56(m_0^U)^2+\frac16m_{\widetilde E_i^c}^2,\\
 m_{\widetilde U_i^c}^2&=\frac53(m_0^U)^2-\frac23m_{\widetilde E_i^c}^2,\\
 m_{\widetilde D_i^c}^2&=\frac53(m_0^U)^2-\frac23m_{\widetilde L_i}^2.
 \label{eq:scalar}
\end{align}
The Higgs-sector flag is chosen such that $\mu$ and $m_A$ are supplied as weak-scale inputs.  This corresponds to a NUHM2-type Higgs parametrization, while the complete setup differs from the conventional NUHM2 because the gaugino, trilinear, and matter-scalar sectors remain nonuniversal and are correlated by the GmSUGRA relations.

The neutralino mass matrix, in the basis $(-i\widetilde B,-i\widetilde W^0,\widetilde H_d^0,\widetilde H_u^0)$, is
\begin{equation}
\resizebox{0.98\columnwidth}{!}{$\displaystyle
{\cal M}_{\widetilde\chi^0}=\begin{pmatrix}
M_1&0&-m_Zs_Wc_\beta&m_Zs_Ws_\beta\\
0&M_2&m_Zc_Wc_\beta&-m_Zc_Ws_\beta\\
-m_Zs_Wc_\beta&m_Zc_Wc_\beta&0&-\mu\\
m_Zs_Ws_\beta&-m_Zc_Ws_\beta&-\mu&0
\end{pmatrix}. $}
\label{eq:neutralino}
\end{equation}
In the region considered below, $|\mu|$ is close to the TeV scale and the non-Higgsino electroweak states are much heavier.  The two light neutral eigenstates therefore form a compressed Higgsino pair.  Their mass difference is taken from the spectrum output after RG evolution and radiative corrections \cite{PierceEtAl1997}.  The numerical scan determines which high-scale soft terms produce the required compression together with the thermal relic abundance.

\textit{Inelastic kinematics and solar capture.---}
For endothermic scattering on a nucleus of mass $m_A$, the minimum incident speed required for a recoil $E_R$ is
\begin{equation}
 v_{\min}(E_R)=\frac{m_AE_R/\mu_{\chi A}+\dmchi}{\sqrt{2m_AE_R}},
 \label{eq:vmin}
\end{equation}
where $\mu_{\chi A}$ is the reduced mass.  A larger $\dmchi$ removes the low-velocity part of the halo distribution and shifts the xenon recoil spectrum toward higher energy \cite{TuckerSmithWeiner2001,TuckerSmithWeiner2005}.  The LZ event consequently selects a narrow sub-MeV range for a TeV Higgsino in standard halo treatments \cite{FreeseTheodosopoulos2026,WuZhangZhu2026,RoddEtAl2026}.

Solar capture reverses the terrestrial advantage of a large splitting.  The solar gravitational potential accelerates the incoming Higgsino, and heavy elements allow inelastic scattering at splittings above the terrestrial xenon reach.  Using the IceCube non-observation of high-energy solar neutrinos, current recasts find a tree-level limit near $506$ keV and a stronger value near $566$ keV when post-capture energy loss and the expected loop-induced channels are included \cite{PospelovRamani2026}.  Related treatments obtain a similar scale \cite{DiMauroShaikh2026,BoseEtAl2026}.  These bounds are derived under assumptions about the local density, velocity distribution, solar composition, capture-annihilation evolution, and annihilation final states.  We therefore use $506$ and $566$ keV as reference boundaries in the figures rather than treating them as model-independent experimental step functions.

\textit{Numerical analysis.---}
The spectrum is generated with ISAJET using two-loop MSSM RG evolution between the weak and unification scales, with radiative electroweak symmetry breaking and a neutralino LSP imposed as in our previous GmSUGRA analyses \cite{BaerEtAlISAJET2000,PaigeEtAlISAJET1998,ISAJET786Manual,KhanEtAlJHEP2026}.  The model considered here is 
so that $\mu$ and $m_A$ are the Higgs-sector inputs, while nonuniversal gaugino, trilinear, and scalar terms are retained.  The ten independent scan parameters are
\begin{equation}
\{m_0,m_{\widetilde E^c},m_{\widetilde L},M_1,M_2,A_t(=A_b),A_\tau,\mu,m_A,\tan\beta\}.
\end{equation}
The derived $M_3$ and squark soft masses are calculated from Eqs.~(\ref{eq:gaugino}) and (\ref{eq:scalar}).

We considered the Higgs bound, B-physics, and current LHC SUSY searches in our results \cite{Aad2012,Chatrchyan2012, Amhis2024,CMSLHCb2023, ATLASStrong2025, ATLASStrong2021, CMSStrong2019, ATLASStop2026, ATLASSUSYSummary2026}.  For the DM discussion, we separate underabundant and overabundant neutralinos from the Planck-compatible interval,
\begin{equation}
0.114\leq \oh\leq0.126,
\label{eq:planck}
\end{equation}
which encompasses the measured cold-DM density once the theory uncertainty used in the scan is included \cite{Planck2018}.

\begin{figure}[t]
\centering
\includegraphics[width=0.98\columnwidth]{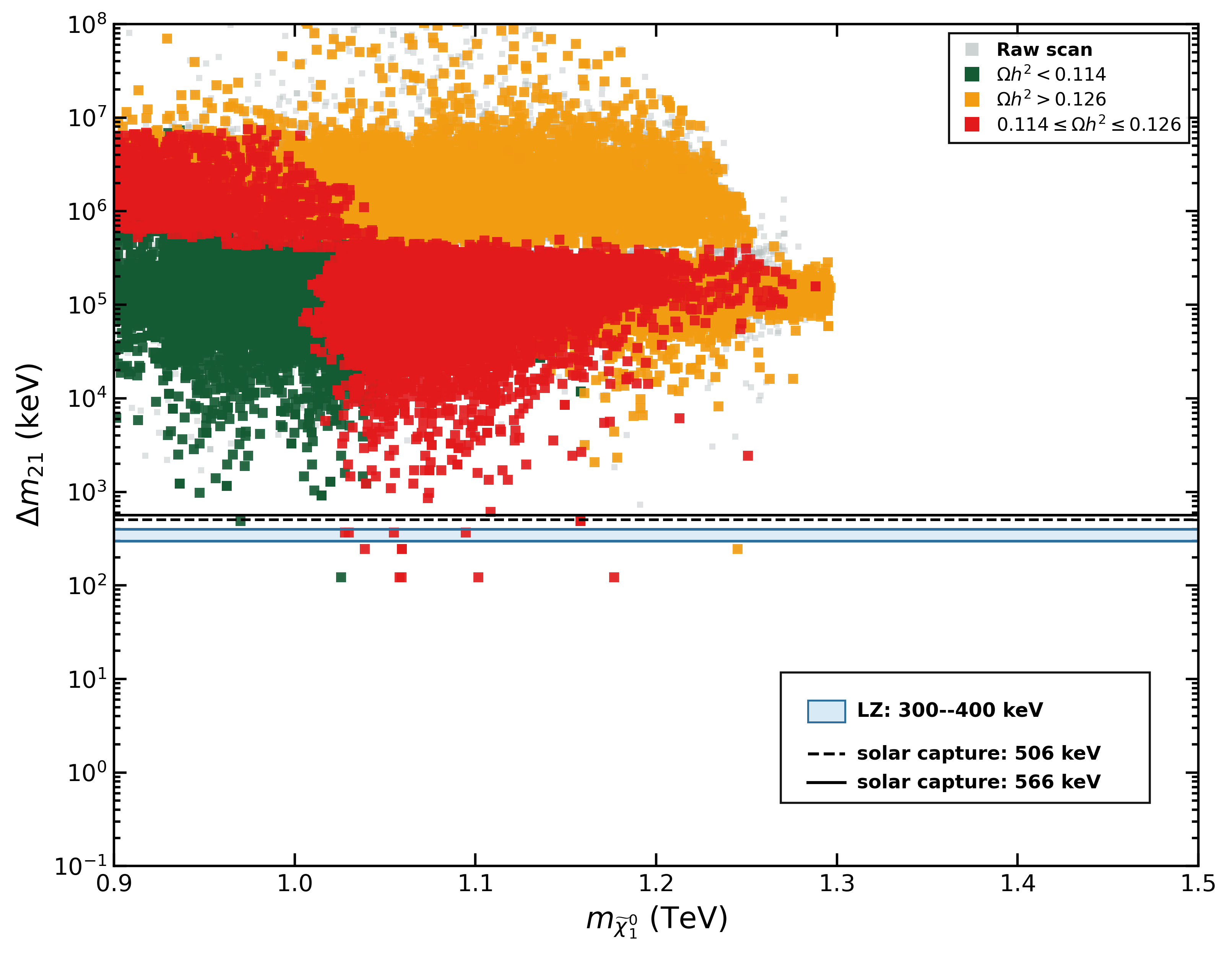}
\caption{Neutral-Higgsino splitting versus $m_{\widetilde\chi_1^0}$ from the complete scan.  Gray points denote the raw sample; green, orange, and red points satisfy the collider/flavor bounds and are classified by the relic density.  The blue band marks $300$--$400$ keV, while the black dashed and solid lines show the $506$ and $566$ keV solar-capture reference values.  The constraint key is placed in the lower-right region where it does not obscure the scan.}
\label{fig:fullscan}
\end{figure}

\textit{Results.---}
Figure~\ref{fig:fullscan} gives the principal result of the scan.  The selected Planck-compatible points form a broad TeV-Higgsino branch.  Most of this branch lies well above the sub-MeV region, while a small extension reaches the few-hundred-keV scale.  The LZ-oriented interval and the solar-capture reference values therefore cut the same GmSUGRA branch at different values of $\dmchi$ rather than defining two unrelated constructions.

\begin{figure}[t]
\centering
\includegraphics[width=0.98\columnwidth]{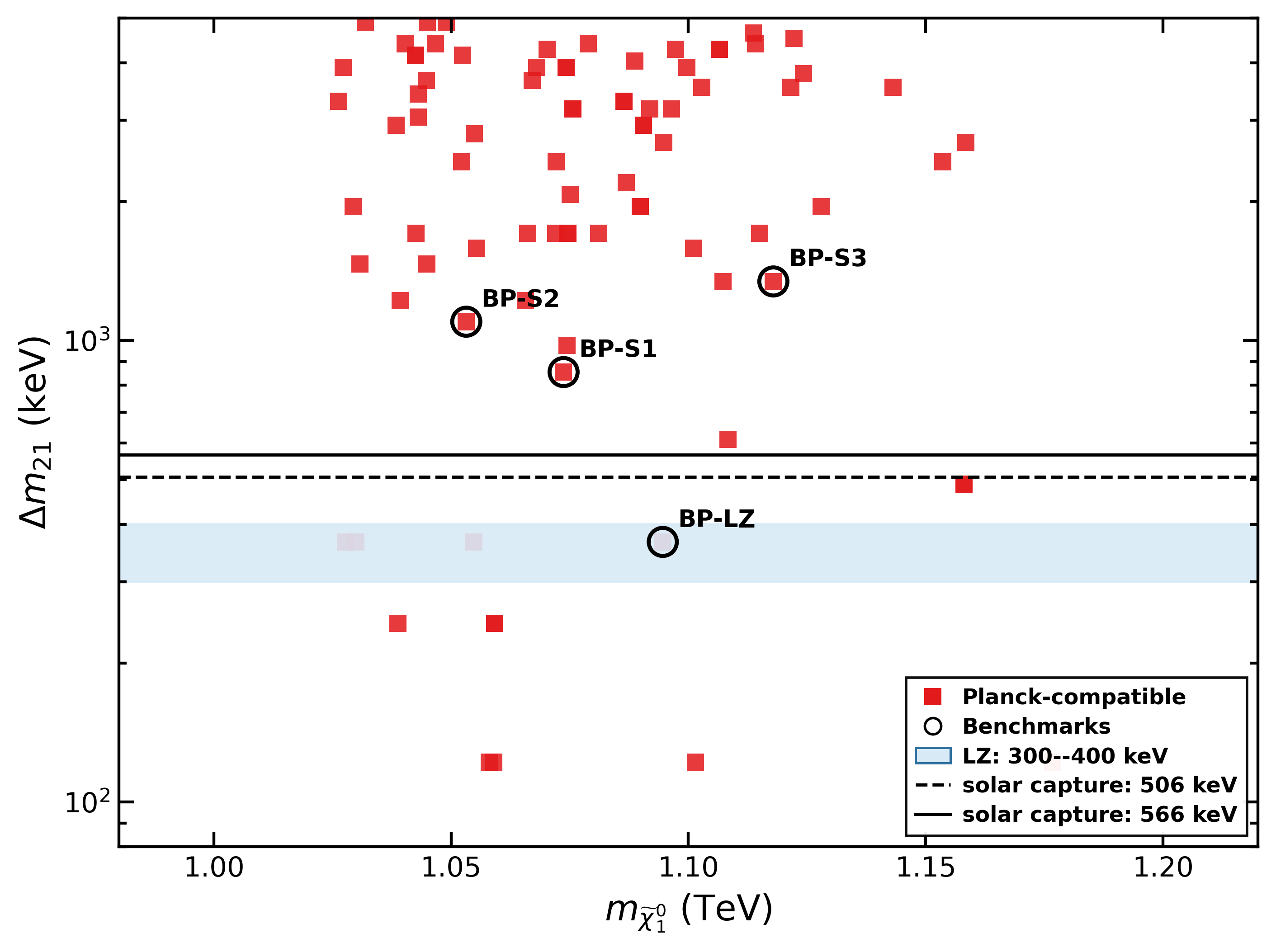}
\caption{Planck-compatible points in the low-splitting region.  The open circles identify the four representative spectra in Tables~\ref{tab:inputs} and \ref{tab:spectrum}.  BP-LZ illustrates the LZ-oriented part of the branch, whereas BP-S1--BP-S3 lie above the current solar-capture reference values.}
\label{fig:zoom}
\end{figure}

The low-splitting region is shown more clearly in Fig.~\ref{fig:zoom}.  A Planck-compatible point with $m_{\widetilde\chi_1^0}=1.095$ TeV and $\dmchi=0.366$ MeV lies in the LZ-oriented interval.  Under the assumptions used in the current solar-capture recasts, this point is below both reference boundaries and cannot be regarded as a viable full-density thermal-Higgsino explanation of the LZ event.  The scan nevertheless continues to Planck-compatible spectra at larger splitting.  The representative points BP-S1, BP-S2, and BP-S3 have $\dmchi=0.854$, $1.099$, and $1.343$ MeV, respectively, and therefore lie above the quoted solar-capture limits.

This distinction is useful because recent work has explored several ways of obtaining a few-hundred-keV Higgsino splitting.  In the MSSM, very heavy electroweak gauginos can suppress the neutral splitting, while a nonuniversal GUT-scale relation can reduce the required gaugino scale through a bino--wino cancellation \cite{DuWang2026}.  Additional neutral-state mixing can produce a similar splitting with TeV-scale gauginos in extended supersymmetric models \cite{BisalCaoLi2026}.  A different possibility is to abandon the thermal $1.1$ TeV Higgsino and move to a much heavier nonthermal state, which changes both the LZ fit and the solar-capture calculation \cite{Langhoff2026}.  Our construction follows a separate route: the Higgsino mass, gaugino sector, and colored spectrum are generated together from the GmSUGRA boundary conditions, and the scan is used to determine which part of that correlated spectrum lies in each experimental regime.

\begin{figure*}[t]
\centering
\includegraphics[width=\textwidth]{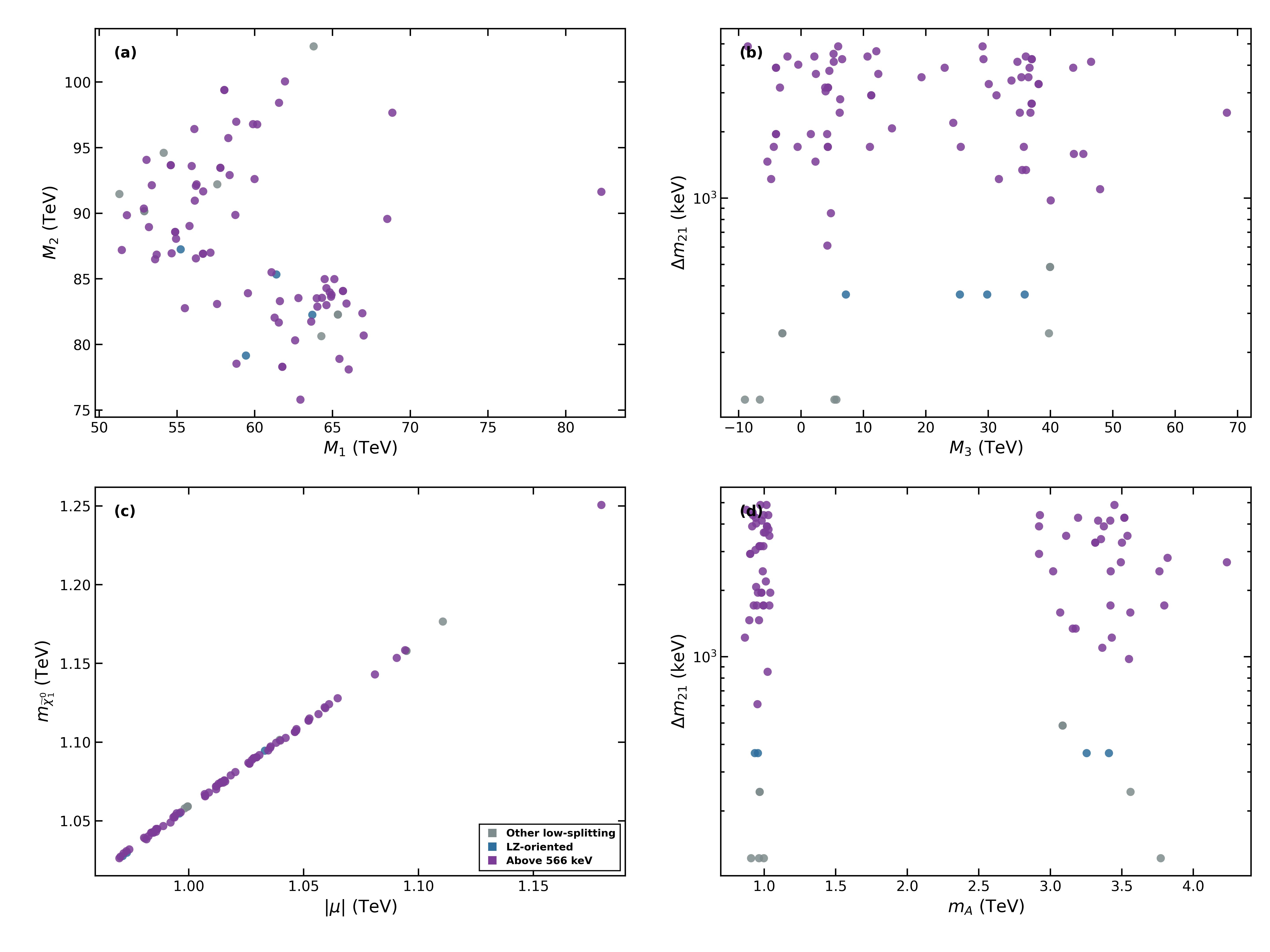}
\caption{GmSUGRA correlations for Planck-compatible points with $80\,{\rm keV}<\dmchi<5$ MeV.  (a) GUT-scale $M_1$--$M_2$ plane; (b) derived $M_3$ versus the neutral splitting; (c) $|\mu|$ versus $m_{\widetilde\chi_1^0}$; (d) $m_A$ versus the neutral splitting.  Blue points fall in the LZ-oriented interval, purple points lie above $566$ keV, and gray points occupy the remaining low-splitting region.}
\label{fig:params}
\end{figure*}

Figure~\ref{fig:params} displays the correlations that are most relevant to this point.  Panel (a) shows the GUT-scale $M_1$--$M_2$ plane.  The LZ-oriented points occupy the same correlated domain as the larger-splitting Planck points; they are not generated by introducing an additional low-energy splitting parameter.  Through Eq.~(\ref{eq:gaugino}), this electroweak-gaugino choice fixes $M_3$, shown against $\dmchi$ in panel (b).  Panel (c) exhibits the nearly linear relation between $|\mu|$ and the lightest neutralino mass, which identifies the TeV state as the Higgsino branch selected by thermal freeze-out.  Panel (d) shows that the sub-MeV and MeV splittings occur for both the lower-$m_A$ and decoupled heavy-Higgs regions sampled by the scan.  Taken together, the four panels show that the experimental selection acts on a correlated high-scale spectrum rather than on an isolated electroweak parameter.

The role of the relic abundance is also important.  A Higgsino with a smaller cosmological fraction could weaken both terrestrial and solar rates after the appropriate density rescaling, but this is not the interpretation pursued here.  The red points in Figs.~\ref{fig:fullscan} and \ref{fig:zoom} satisfy Eq.~(\ref{eq:planck}) and are treated as the dominant DM component.  The survival of the larger-splitting branch is therefore not obtained by lowering the Higgsino abundance.  Instead, it follows directly from the range of splittings generated by the GmSUGRA spectrum.

\begin{figure*}[t]
\centering
\includegraphics[width=\textwidth]{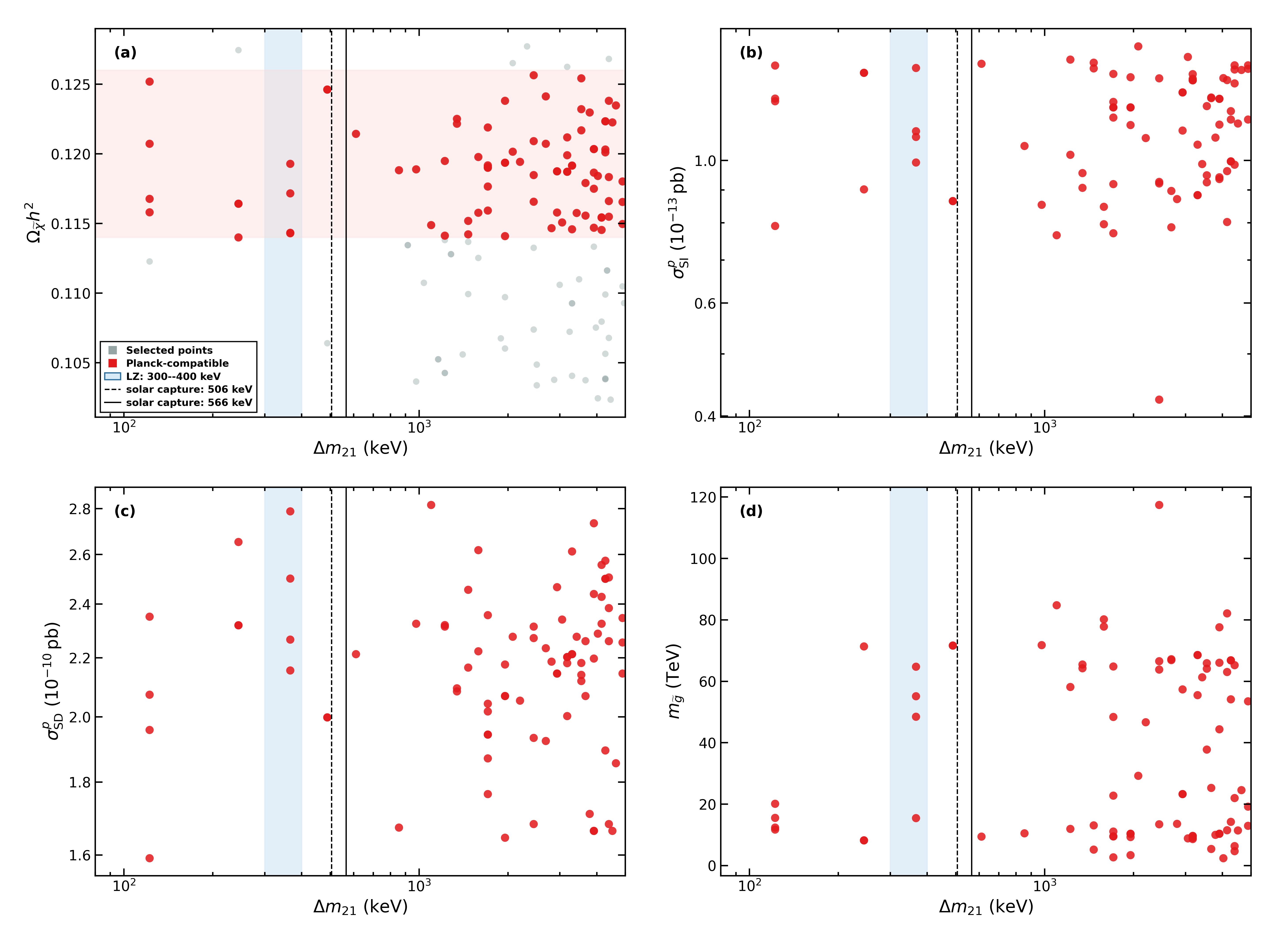}
\caption{Additional observables for the low-splitting scan.  (a) Relic density versus $\dmchi$, with the Planck interval indicated by the horizontal band; (b) spin-independent and (c) spin-dependent elastic proton cross sections; (d) gluino mass.  The blue vertical band marks the LZ-oriented interval and the dashed/solid lines denote $506/566$ keV.}
\label{fig:obs}
\end{figure*}

Figure~\ref{fig:obs} displays observables that are secondary to the principal mass-splitting analysis but useful for characterizing the selected spectra.  The relic-density panel shows that the Planck-compatible solutions occur both in the LZ-oriented interval and above the solar-capture reference values.  The elastic proton cross sections remain small throughout the same branch.  These elastic quantities are not the interaction responsible for the inelastic solar-capture constraint, which is driven by the off-diagonal Higgsino $Z$ coupling.  The gluino panel illustrates the correlated colored spectrum implied by the GmSUGRA gaugino relation.

The elastic spin-independent and spin-dependent cross sections of the representative region are small, as summarized in Table~\ref{tab:spectrum}.  This is consistent with the strongly Higgsino-dominated LSP and heavy non-Higgsino spectrum.  These small elastic rates do not remove the solar-capture constraint in the LZ-sized region because the relevant capture mechanism is the inelastic $Z$-mediated transition.  The distinction between elastic direct-detection observables and inelastic capture is therefore kept explicit throughout the analysis.

A numerical point deserves emphasis.  The archived spectrum chain stores the pole masses entering the appended mass-difference field with single-precision variables.  Around a TeV this limits the useful numerical resolution of the dedicated sub-MeV splitting output to roughly $\mathcal O(0.1)$ MeV.  We therefore use the low-splitting values to identify the physical regions and choose the solar-compatible benchmarks well above the $566$ keV reference line.  No conclusion in the paper relies on resolving a difference of a few tens of keV near that boundary.

\textit{Conclusion.---}
We have studied thermal Higgsino dark matter in the MSSM with GmSUGRA boundary conditions using the complete scan that generated the compressed neutralino spectrum.  The high-scale gaugino and scalar relations correlate the TeV Higgsino with the rest of the supersymmetric spectrum, while the NUHM2-type Higgs parametrization allows $\mu$ and $m_A$ to control the electroweak Higgs sector directly.

The scan contains a Planck-compatible Higgsino branch near the thermal mass with neutral splittings ranging from the few-hundred-keV region to the MeV scale.  The part with $\dmchi\simeq0.3$--$0.4$ MeV overlaps the kinematic interval relevant to the LZ high-recoil event.  Current solar-capture reinterpretations of the IceCube null result disfavor that part of the branch for a full-density thermal Higgsino.  The same GmSUGRA construction, however, yields nearby Planck-compatible spectra with larger splittings, including the representative points at $0.85$--$1.34$ MeV discussed above.  These points lie beyond the quoted solar-capture limits while preserving the TeV thermal Higgsino and the correlated high-scale spectrum.

The combined LZ and solar-neutrino information therefore acts as a selection on the GmSUGRA Higgsino branch.  The few-hundred-keV solutions remain useful for identifying the high-scale origin of the LZ-motivated compression, whereas the larger-splitting solutions describe the part of the thermal branch compatible with the current solar-capture constraints.  A dedicated solar-capture calculation using the GmSUGRA mixing matrices and annihilation fractions would provide the next step beyond the reference-limit treatment adopted here.

{\textbf{Acknowledgements.}---}
TL is supported in part by the National Key Research and Development Program of China Grant No. 2020YFC2201504, by the Projects No. 11875062, No. 11947302, No. 12047503, and No. 12275333 supported by the National Natural Science Foundation of China, by the Key Research Program of the Chinese Academy of Sciences, Grant No. XDPB15, by the Scientific Instrument Developing Project of the Chinese Academy of Sciences, Grant No. YJKYYQ20190049, by the International Partnership Program of Chinese Academy of Sciences for Grand Challenges, Grant No. 112311KYSB20210012, and by the Henan Province Outstanding Foreign Scientist Studio Project, No.GZS2025008. This work was supported by the High Performance Computing Platform of Henan Normal University. The authors also acknowledge the use of ChatGPT for assistance in improving the English language and clarity of the manuscript.

\begin{table*}[t]
\caption{Independent GmSUGRA inputs and derived GUT-scale soft masses for the representative spectra.  Dimensionful quantities are in GeV.  BP-LZ lies in the LZ-oriented interval; BP-S1--BP-S3 lie above the stronger solar-capture reference value.}
\label{tab:inputs}
\scriptsize
\setlength{\tabcolsep}{4pt}
\begin{ruledtabular}
\begin{tabular}{lrrrr}
Parameter & BP-LZ & BP-S1 & BP-S2 & BP-S3\\
\hline
$m_0$ & 65779 & 65597 & 42562 & 45414\\
$m_{\widetilde E^c}$ & 23940 & 24455 & 28464 & 25467\\
$m_{\widetilde L}$ & 26466 & 27042 & 31027 & 27706\\
$M_1$ & 61379 & 61942 & 66040 & 64811\\
$M_2$ & 85335 & 100050 & 78100 & 83976\\
$M_3$ & 25446 & 4785 & 47951 & 36064\\
$A_t=A_b$ & 6419.8 & 6617.9 & 10326 & 11762\\
$A_\tau$ & -22131 & -24188 & -27169 & -28875\\
$\mu$ & 1033.1 & 1012.9 & 993.9 & 1056.4\\
$m_A$ & 933.61 & 1022.3 & 3362.6 & 3175.8\\
$\tan\beta$ & 21.1 & 22.1 & 22.3 & 21.6\\
$m_{\widetilde Q}$ & 60838 & 60708 & 40554 & 42741\\
$m_{\widetilde D^c}$ & 82125 & 81756 & 48759 & 54089\\
$m_{\widetilde U^c}$ & 82640 & 82298 & 49790 & 54818\\
\end{tabular}
\end{ruledtabular}
\end{table*}

\begin{table*}[t]
\caption{Weak-scale spectrum and selected observables for the benchmark points.  Masses are in GeV unless another unit is shown.  The neutral splitting is taken from the dedicated mass-difference field in the scan output.}
\label{tab:spectrum}
\scriptsize
\setlength{\tabcolsep}{4pt}
\begin{ruledtabular}
\begin{tabular}{lrrrr}
Observable & BP-LZ & BP-S1 & BP-S2 & BP-S3\\
\hline
$m_h$ & 127.80 & 127.97 & 127.93 & 127.77\\
$m_H$ & 939.8 & 1029.1 & 3384.7 & 3196.7\\
$m_A$ & 933.61 & 1022.3 & 3362.6 & 3175.8\\
$m_{H^\pm}$ & 937.06 & 1025.5 & 3363.5 & 3176.8\\
$m_{\widetilde\chi_1^0}$ & 1094.6 & 1073.7 & 1053.2 & 1117.9\\
$m_{\widetilde\chi_2^0}$ & 1094.6 & 1073.7 & 1053.2 & 1117.9\\
$m_{\widetilde\chi_3^0}$ & 30844 & 31283 & 33072 & 32487\\
$m_{\widetilde\chi_4^0}$ & 72776 & 85188 & 66597 & 71412\\
$m_{\widetilde\chi_1^\pm}$ & 1150.0 & 1127.5 & 1107.2 & 1173.8\\
$m_{\widetilde\chi_2^\pm}$ & 72275 & 84832 & 65962 & 70891\\
$m_{\widetilde g}$ & 48524 & 10558 & 84786 & 65471\\
$m_{\widetilde u_L},m_{\widetilde u_R}$ & 86086, 90587 & 83968, 82914 & 90429, 85532 & 82233, 76592\\
$m_{\widetilde d_L},m_{\widetilde d_R}$ & 86086, 90750 & 83968, 83114 & 90429, 83831 & 82233, 75360\\
$m_{\widetilde t_1},m_{\widetilde t_2}$ & 75200, 88599 & 74157, 81413 & 82172, 82269 & 73353, 74446\\
$m_{\widetilde b_1},m_{\widetilde b_2}$ & 69080, 75596 & 60873, 74577 & 66786, 82671 & 58625, 74770\\
$m_{\widetilde\nu_e}$ & 56123 & 63969 & 56257 & 57064\\
$m_{\widetilde\nu_\tau}$ & 55484 & 63406 & 55493 & 56408\\
$m_{\widetilde e_L},m_{\widetilde e_R}$ & 56131, 36121 & 63971, 36760 & 56290, 36699 & 57078, 35097\\
$m_{\widetilde\tau_1},m_{\widetilde\tau_2}$ & 34634, 55593 & 34949, 63411 & 34832, 55589 & 33248, 56447\\
$\dmchi$ [keV] & 366.21 & 854.49 & 1098.6 & 1342.8\\
$\oh$ & 0.11929 & 0.11884 & 0.11490 & 0.12252\\
${\rm BR}(b\to s\gamma)$ & $3.421\times10^{-4}$ & $3.377\times10^{-4}$ & $3.099\times10^{-4}$ & $3.106\times10^{-4}$\\
${\rm BR}(B_s\to\mu^+\mu^-)$ & $3.844\times10^{-9}$ & $3.843\times10^{-9}$ & $3.838\times10^{-9}$ & $3.838\times10^{-9}$\\
$\sigma_{\rm SI}^{p}$ [pb] & $1.09\times10^{-13}$ & $1.05\times10^{-13}$ & $7.65\times10^{-14}$ & $9.07\times10^{-14}$\\
$\sigma_{\rm SD}^{p}$ [pb] & $2.16\times10^{-10}$ & $1.67\times10^{-10}$ & $2.82\times10^{-10}$ & $2.09\times10^{-10}$\\
\end{tabular}
\end{ruledtabular}
\end{table*}

\bibliography{references_fixed_APS}

@misc{LZ2026HighRecoil,
  author = {{LZ Collaboration}},
  title = {Search for dark matter particle interactions in an extended nuclear recoil energy window with the LUX-ZEPLIN (LZ) experiment},
  year = {2026},
  eprint = {2609.02823},
  archivePrefix = {arXiv},
  primaryClass = {hep-ex}
}

@misc{FreeseTheodosopoulos2026,
  author = {Katherine Freese and Dionysios P. Theodosopoulos},
  title = {Higgsino Dark Matter Interpretation of the LUX-ZEPLIN 248 keV Nuclear-Recoil Event},
  year = {2026},
  eprint = {2609.01583},
  archivePrefix = {arXiv},
  primaryClass = {hep-ph}
}

@misc{WuZhangZhu2026,
  author = {Lei Wu and Yang Zhang and Bin Zhu},
  title = {TeV Higgsino Dark Matter from LZ Nuclear Recoil to Fermi-LAT Gamma Rays},
  year = {2026},
  eprint = {2609.01590},
  archivePrefix = {arXiv},
  primaryClass = {hep-ph}
}

@misc{FanReece2026,
  author = {JiJi Fan and Matthew Reece},
  title = {Higgsino Above the Sea of Fog},
  year = {2026},
  eprint = {2609.01504},
  archivePrefix = {arXiv},
  primaryClass = {hep-ph}
}

@misc{Yin2026,
  author = {Wen Yin},
  title = {A PQ-Symmetric High-Scale SUSY Interpretation of the LZ High-Energy Recoil},
  year = {2026},
  eprint = {2609.01892},
  archivePrefix = {arXiv},
  primaryClass = {hep-ph}
}

@misc{DuWang2026,
  author = {Xiaokang Du and Fei Wang},
  title = {TeV Higgsino Interpretation of the LZ High-Recoil Event with Intermediate-Scale Electroweak Gauginos},
  year = {2026},
  eprint = {2609.04163},
  archivePrefix = {arXiv},
  primaryClass = {hep-ph}
}

@misc{RoddEtAl2026,
  author = {Nicholas L. Rodd and Benjamin R. Safdi and Tracy R. Slatyer and Weishuang Linda Xu},
  title = {Confronting the Higgsino Interpretation of the LZ Event with the High-Energy Sideband},
  year = {2026},
  eprint = {2609.04175},
  archivePrefix = {arXiv},
  primaryClass = {hep-ph}
}

@misc{BisalCaoLi2026,
  author = {Subhadip Bisal and Junjie Cao and Fei Li},
  title = {Higgsino Dark Matter Interpretation of the LZ High-Recoil Event in the GNMSSM with TeV-Scale Gauginos},
  year = {2026},
  eprint = {2609.07811},
  archivePrefix = {arXiv},
  primaryClass = {hep-ph}
}

@misc{BoseEtAl2026,
  author = {Debajit Bose and Akash Kumar Saha and Deep Jyoti Das and Rinchen Sherpa and Abhijeet Singh and Durba Ghosh and Jaya Doliya and Subhadip Bouri and Biprajit Mondal and Ranjini Mondol and Nirmal Raj and Ranjan Laha},
  title = {Not so good neutrinos for Higgsino dark matter as LZ excess: stringent limits from Super-Kamiokande and IceCube},
  year = {2026},
  eprint = {2609.07807},
  archivePrefix = {arXiv},
  primaryClass = {hep-ph}
}

@article{TuckerSmithWeiner2001,
  author = {David Tucker-Smith and Neal Weiner},
  title = {Inelastic dark matter},
  journal = {Phys. Rev. D},
  volume = {64},
  pages = {043502},
  year = {2001},
  eprint = {hep-ph/0101138},
  doi = {10.1103/PhysRevD.64.043502}
}

@article{TuckerSmithWeiner2005,
  author = {David Tucker-Smith and Neal Weiner},
  title = {The Status of inelastic dark matter},
  journal = {Phys. Rev. D},
  volume = {72},
  pages = {063509},
  year = {2005},
  eprint = {hep-ph/0402065},
  doi = {10.1103/PhysRevD.72.063509}
}

@article{MartinPrimer,
  author = {Stephen P. Martin},
  title = {A Supersymmetry primer},
  journal = {Adv. Ser. Direct. High Energy Phys.},
  volume = {18},
  pages = {1--98},
  year = {1998},
  eprint = {hep-ph/9709356},
  doi = {10.1142/9789812839657_0001}
}

@article{HaberKane1985,
  author = {Howard E. Haber and Gordon L. Kane},
  title = {The Search for Supersymmetry: Probing Physics Beyond the Standard Model},
  journal = {Phys. Rept.},
  volume = {117},
  pages = {75--263},
  year = {1985},
  doi = {10.1016/0370-1573(85)90051-1}
}

@article{JungmanEtAl1996,
  author = {Gerard Jungman and Marc Kamionkowski and Kim Griest},
  title = {Supersymmetric dark matter},
  journal = {Phys. Rept.},
  volume = {267},
  pages = {195--373},
  year = {1996},
  eprint = {hep-ph/9506380},
  doi = {10.1016/0370-1573(95)00058-5}
}

@article{LiNanopoulos2010,
  author = {Tianjun Li and Dimitri V. Nanopoulos},
  title = {Generalizing Minimal Supergravity},
  journal = {Phys. Lett. B},
  volume = {692},
  pages = {121--125},
  year = {2010},
  eprint = {1002.4183},
  archivePrefix = {arXiv},
  primaryClass = {hep-ph},
  doi = {10.1016/j.physletb.2010.07.025}
}

@article{BalazsEtAl2010,
  author = {Csaba Balazs and Tianjun Li and Dimitri V. Nanopoulos and Fei Wang},
  title = {Supersymmetry Breaking Scalar Masses and Trilinear Soft Terms in Generalized Minimal Supergravity},
  journal = {JHEP},
  volume = {2010},
  number = {09},
  pages = {003},
  year = {2010},
  eprint = {1006.5559},
  archivePrefix = {arXiv},
  primaryClass = {hep-ph},
  doi = {10.1007/JHEP09(2010)003}
}

@article{LiRaza2015,
  author = {Tianjun Li and Shabbar Raza},
  title = {Electroweak supersymmetry from the generalized minimal supergravity model in the MSSM},
  journal = {Phys. Rev. D},
  volume = {91},
  pages = {055016},
  year = {2015},
  eprint = {1409.3930},
  archivePrefix = {arXiv},
  primaryClass = {hep-ph},
  doi = {10.1103/PhysRevD.91.055016}
}

@article{KhanEtAl2024,
  author = {Imtiaz Khan and Waqas Ahmed and Tianjun Li and Shabbar Raza},
  title = {Right-handed slepton bulk regions for dark matter in a generalized minimal supergravity},
  journal = {Phys. Rev. D},
  volume = {109},
  pages = {075051},
  year = {2024},
  eprint = {2312.07863},
  archivePrefix = {arXiv},
  primaryClass = {hep-ph},
  doi = {10.1103/PhysRevD.109.075051}
}

@article{KhanEtAl2025Light,
  author = {Imtiaz Khan and Waqas Ahmed and Tianjun Li and Shabbar Raza and Ali Muhammad},
  title = {The Light Neutralino Dark Matter in the Generalized Minimal Supergravity (GmSUGRA)},
  journal = {Phys. Lett. B},
  volume = {869},
  pages = {139844},
  year = {2025},
  eprint = {2501.12039},
  archivePrefix = {arXiv},
  primaryClass = {hep-ph},
  doi = {10.1016/j.physletb.2025.139844}
}

@article{KhanEtAl2026EWSUSY,
  author = {Imtiaz Khan and Ali Muhammad and Tianjun Li and Shabbar Raza},
  title = {Revisiting the Electroweak Supersymmetry from the Generalized Minimal Supergravity},
  journal = {Phys. Rev. D},
  volume = {114},
  pages = {035006},
  year = {2026},
  eprint = {2506.18442},
  archivePrefix = {arXiv},
  primaryClass = {hep-ph},
  doi = {10.1103/m2nv-kshl}
}

@article{KhanEtAlJHEP2026,
  author = {Imtiaz Khan and Ali Muhammad and Tianjun Li and Shabbar Raza and Pirzada Waqar and Mussawir Khan},
  title = {The light neutralino dark matter at Future Colliders in the MSSM with the generalized minimal supergravity (GmSUGRA)},
  journal = {JHEP},
  volume = {2026},
  number = {06},
  pages = {115},
  year = {2026},
  eprint = {2509.23356},
  archivePrefix = {arXiv},
  primaryClass = {hep-ph},
  doi = {10.1007/JHEP06(2026)115}
}

@misc{BaerEtAlISAJET2000,
  author = {Howard Baer and Frank E. Paige and Serban D. Protopopescu and Xerxes Tata},
  title = {ISAJET 7.48: A Monte Carlo event generator for pp, anti-p p, and e+e- reactions},
  year = {2000},
  eprint = {hep-ph/0001086},
  archivePrefix = {arXiv}
}

@misc{PaigeEtAlISAJET1998,
  author = {Frank E. Paige and Serban D. Protopopescu and Howard Baer and Xerxes Tata},
  title = {ISAJET 7.37: A Monte Carlo event generator for pp, anti-p p, and e+e- reactions},
  year = {1998},
  eprint = {hep-ph/9804321},
  archivePrefix = {arXiv}
}

@article{PierceEtAl1997,
  author = {Damien M. Pierce and Jonathan A. Bagger and Konstantin T. Matchev and Ren-Jie Zhang},
  title = {Precision corrections in the minimal supersymmetric standard model},
  journal = {Nucl. Phys. B},
  volume = {491},
  pages = {3--67},
  year = {1997},
  eprint = {hep-ph/9606211},
  doi = {10.1016/S0550-3213(96)00683-9}
}

@article{Planck2018,
  author = {{Planck Collaboration}},
  title = {Planck 2018 results. VI. Cosmological parameters},
  journal = {Astron. Astrophys.},
  volume = {641},
  pages = {A6},
  year = {2020},
  eprint = {1807.06209},
  archivePrefix = {arXiv},
  primaryClass = {astro-ph.CO},
  doi = {10.1051/0004-6361/201833910}
}

@misc{ISAJET786Manual,
  author = {H. Baer and F. E. Paige and S. D. Protopopescu and X. Tata},
  title = {ISAJET 7.86: A Monte Carlo event generator for pp, anti-p p, and e+e- reactions},
  year = {2016},
  note = {ISAJET manual; the NUHM option accepts weak-scale $\mu$ and $m_A$ as nonuniversal Higgs inputs}
}

@misc{IceCubeSolar2025,
  author = {{IceCube Collaboration}},
  title = {Search for High-Energy Neutrinos From the Sun Using Ten Years of IceCube Data},
  year = {2025},
  eprint = {2507.08457},
  archivePrefix = {arXiv},
  primaryClass = {hep-ex}
}

@misc{PospelovRamani2026,
  author = {Maxim Pospelov and Harikrishnan Ramani},
  title = {Strong Constraints on Higgsino Dark Matter from Solar Capture},
  year = {2026},
  eprint = {2609.02775},
  archivePrefix = {arXiv},
  primaryClass = {hep-ph}
}

@misc{DiMauroShaikh2026,
  author = {Mattia Di Mauro and Halim Shaikh},
  title = {Solar Capture Tests of Inelastic Dark Matter after the LZ High-Recoil Event},
  year = {2026},
  eprint = {2609.06760},
  archivePrefix = {arXiv},
  primaryClass = {hep-ph}
}

@misc{Langhoff2026,
  author = {Kevin Langhoff},
  title = {Heavy Higgsino Interpretation of the LZ Event},
  year = {2026},
  eprint = {2609.09385},
  archivePrefix = {arXiv},
  primaryClass = {hep-ph}
}

@misc{GhoshEtAl2026,
  author = {Aditya Ghosh and Ilumi Chavez and Chris Kelso},
  title = {Confronting the Higgsino Interpretation of the LZ Event with Astrophysical Uncertainties and the Solar Capture Constraints},
  year = {2026},
  eprint = {2609.15321},
  archivePrefix = {arXiv},
  primaryClass = {hep-ph}
}

@article{Aad2012,
  author = {{ATLAS Collaboration}},
  title = {Observation of a new particle in the search for the Standard Model Higgs boson with the ATLAS detector at the LHC},
  journal = {Phys. Lett. B},
  volume = {716},
  pages = {1--29},
  year = {2012},
  eprint = {1207.7214},
  archivePrefix = {arXiv},
  primaryClass = {hep-ex},
  doi = {10.1016/j.physletb.2012.08.020}
}

@article{Chatrchyan2012,
  author = {{CMS Collaboration}},
  title = {Observation of a new boson at a mass of 125 GeV with the CMS experiment at the LHC},
  journal = {Phys. Lett. B},
  volume = {716},
  pages = {30--61},
  year = {2012},
  eprint = {1207.7235},
  archivePrefix = {arXiv},
  primaryClass = {hep-ex},
  doi = {10.1016/j.physletb.2012.08.021}
}

@article{Amhis2024,
  author = {Yasmine S. Amhis and others},
  collaboration = {HFLAV},
  title = {Averages of $b$-hadron, $c$-hadron, and $\tau$-lepton properties as of 2023},
  journal = {Phys. Rev. D},
  volume = {110},
  pages = {030001},
  year = {2024},
  eprint = {2411.18639},
  archivePrefix = {arXiv},
  primaryClass = {hep-ex}
}

@article{CMSLHCb2023,
  author = {{CMS Collaboration} and {LHCb Collaboration}},
  title = {Observation of the rare $B_s^0\to\mu^+\mu^-$ decay from combined CMS and LHCb data},
  journal = {Nature},
  volume = {522},
  pages = {68--72},
  year = {2015},
  doi = {10.1038/nature14474}
}

@article{ATLASStrong2025,
  author = {{ATLAS Collaboration}},
  title = {Search for squarks and gluinos in $pp$ collisions at
  $\sqrt{s}=13$ TeV and $13.6$ TeV in events with
  $\tau$-leptons, jets and missing transverse momentum using
  the ATLAS detector},
  journal = {Eur. Phys. J. C},
  volume = {85},
  pages = {1437},
  year = {2025},
  eprint = {2507.00296},
  archivePrefix = {arXiv},
  primaryClass = {hep-ex},
  doi = {10.1140/epjc/s10052-025-14957-6}
}

@article{ATLASStrong2021,
  author = {{ATLAS Collaboration}},
  title = {Search for squarks and gluinos in final states with jets
  and missing transverse momentum using 139 fb$^{-1}$ of
  $\sqrt{s}=13$ TeV $pp$ collision data with the ATLAS detector},
  journal = {JHEP},
  volume = {2021},
  number = {02},
  pages = {143},
  year = {2021},
  eprint = {2010.14293},
  archivePrefix = {arXiv},
  primaryClass = {hep-ex},
  doi = {10.1007/JHEP02(2021)143}
}

@article{CMSStrong2019,
  author = {{CMS Collaboration}},
  title = {Search for supersymmetry in proton-proton collisions at
  13 TeV in final states with jets and missing transverse momentum},
  journal = {JHEP},
  volume = {2019},
  number = {10},
  pages = {244},
  year = {2019},
  eprint = {1908.04722},
  archivePrefix = {arXiv},
  primaryClass = {hep-ex},
  doi = {10.1007/JHEP10(2019)244}
}

@misc{ATLASStop2026,
  author = {{ATLAS Collaboration}},
  title = {Search for direct pair production of top squarks in
  $pp$ collisions at $\sqrt{s}=13$ TeV and $13.6$ TeV in events
  with two oppositely charged leptons using the ATLAS detector},
  year = {2026},
  eprint = {2603.16191},
  archivePrefix = {arXiv},
  primaryClass = {hep-ex},
  reportNumber = {CERN-EP-2026-063}
}

@article{Khan:2026nwp,
    author = "Khan, Imtiaz and Capozziello, Salvatore and Mustafa, G. and Atamurotov, Farruh and Abdujabbarov, Ahmadjon and Yuan, Chengxun",
    title = "{Nuclear interference versus dark sector excitation in the 248 keV LUX-ZEPLIN recoil candidate}",
    journal = "arXiv e-prints",
    eprint = "2609.09230",
    archivePrefix = "arXiv",
    primaryClass = "hep-ph",
    year = "2026"
}

@techreport{ATLASSUSYSummary2026,
  author      = {{ATLAS Collaboration}},
  title       = {SUSY March 2026 Summary Plot Update},
  institution = {CERN},
  number      = {ATL-PHYS-PUB-2026-003},
  year        = {2026},
  month       = mar,
  note        = {ATLAS public note, 18 March 2026}
}

@article{Khan:2023ryc,
    author = "Khan, Imtiaz and Ahmed, Waqas and Li, Tianjun and Raza, Shabbar",
    title = "{Right-handed slepton bulk regions for dark matter in a generalized minimal supergravity}",
    eprint = "2312.07863",
    archivePrefix = "arXiv",
    primaryClass = "hep-ph",
    doi = "10.1103/PhysRevD.109.075051",
    journal = "Phys. Rev. D",
    volume = "109",
    number = "7",
    pages = "075051",
    year = "2024"
}

@article{AhmedEtAl2018Natural,
  author        = {Ahmed, Waqas and Bi, Xiao-Jun and Li, Tianjun and Niu, Jia Shu and Raza, Shabbar and Xiang, Qian-Fei and Yin, Peng-Fei},
  title         = {Status of Natural Supersymmetry from the GmSUGRA in Light of the Current LHC Run-2 and LUX Data},
  journal       = {Phys. Rev. D},
  volume        = {98},
  number        = {1},
  pages         = {015040},
  year          = {2018},
  eprint        = {1709.06371},
  archivePrefix = {arXiv},
  primaryClass  = {hep-ph}
}

@article{AhmedEtAl2022MuonG2,
  author        = {Ahmed, Waqas and Khan, Imtiaz and Li, Jinmian and Li, Tianjun and Raza, Shabbar and Zhang, Wenxing},
  title         = {The Natural Explanation of the Muon Anomalous Magnetic Moment via the Electroweak Supersymmetry from the GmSUGRA in the MSSM},
  journal       = {Phys. Lett. B},
  volume        = {827},
  pages         = {136879},
  year          = {2022},
  eprint        = {2104.03491},
  archivePrefix = {arXiv},
  primaryClass  = {hep-ph}
}

@misc{OkadaSeto2026,
  author        = {Okada, Nobuchika and Seto, Osamu},
  title         = {Inelastic $B-L$ Scalar Dark Matter and the LUX-ZEPLIN Event},
  year          = {2026},
  eprint        = {2609.06909},
  archivePrefix = {arXiv},
  primaryClass  = {hep-ph}
}

@misc{SmirnovEtAl2026,
  author        = {Smirnov, Juri and Griffith, Spencer and Beacom, John F.},
  title         = {Inelastic Signatures of Electroweak Dark Matter},
  year          = {2026},
  eprint        = {2609.04144},
  archivePrefix = {arXiv},
  primaryClass  = {hep-ph}
}

@misc{YuanEtAl2026,
  author        = {Yuan, Guan-Wen and Zhang, Bo and Cao, Wen-Yu and Feng, Lei and Yang, Ruizhi},
  title         = {ALP-mediated Inelastic Dark Matter and the LUX-ZEPLIN High-Recoil Candidate Event LZ230616},
  year          = {2026},
  eprint        = {2609.08893},
  archivePrefix = {arXiv},
  primaryClass  = {hep-ph}
}

@misc{FanHeWangZhao2026,
  author        = {Fan, Zi-Tong and He, Hong-Jian and Wang, Yu-Chen and Zhao, Yue},
  title         = {Inelastic Dark Matter and High-Energy Recoil Signatures in LZ},
  year          = {2026},
  eprint        = {2609.10491},
  archivePrefix = {arXiv},
  primaryClass  = {hep-ph}
}

@misc{Yamashita2026,
  author        = {Yamashita, Kimiko},
  title         = {Inelastic Dark Photon Dark Matter for the LUX-ZEPLIN High-Recoil Event and the Galactic Halo Gamma-Ray Excess},
  year          = {2026},
  eprint        = {2609.02868},
  archivePrefix = {arXiv},
  primaryClass  = {hep-ph}
}

@misc{AhmedLeontarisDarkDimension2026,
  author        = {Ahmed, Waqas and Leontaris, George K.},
  title         = {A Dark-Dimension Origin of Geometric Inelastic Dark Matter: The LUX-ZEPLIN High-Recoil Event and Multi-Target Tests},
  year          = {2026},
  eprint        = {2609.07138},
  archivePrefix = {arXiv},
  primaryClass  = {hep-ph}
}

@misc{AhmedAhmadRehman2026,
  author        = {Ahmed, Waqas and Ahmad, Ammara and Rehman, Mansoor Ur},
  title         = {Xenon Isotope Filtering at the Kinematic Edge of Inelastic Dark Matter},
  year          = {2026},
  eprint        = {2609.15634},
  archivePrefix = {arXiv},
  primaryClass  = {hep-ph}
}

@misc{AhmedLeontarisKK2026,
  author        = {Ahmed, Waqas and Leontaris, George K.},
  title         = {Kaluza--Klein Dark-Photon Mediation of Inelastic Dark Matter at LUX-ZEPLIN},
  year          = {2026},
  eprint        = {2609.22739},
  archivePrefix = {arXiv},
  primaryClass  = {hep-ph}
}
\end{document}